\documentclass[table,xcdraw,smallextended]{svjour3}
\usepackage[utf8]{inputenc}
\usepackage[anythingbreaks]{breakurl}
\usepackage{url} % Breaking long urls
\usepackage{pseudocode}
\usepackage[linesnumbered,boxed]{algorithm2e} % To have nice algorithms
\usepackage{booktabs} % Tables
\usepackage{makecell} % Break line inside cells
\usepackage{lscape} % Landscape tables
\usepackage{graphicx}
\usepackage{tcolorbox}
\usepackage{hyperref} % Auto references with deduced names
\usepackage{numprint} % Number format
\usepackage{listings} % Code listings
\usepackage{csquotes} % Quotes
\usepackage{amssymb} % Definitions
\usepackage{paralist} % To have lists inlined the text
\usepackage{multirow} % Make cells in tabulars span multiple rows
\usepackage{adjustbox} % Adjust the content to a given size
\usepackage{tablefootnote} % Footnotes in tables
\usepackage{float}

\newcommand{\theoriginalauthors}{Niedermayr et al. \cite{niedermayr_will_2016}}

\newcommand{\pseudotested}{pseudo-tested}

\newcommand{\pit}{PITest}
\newcommand{\java}{Java}
\newcommand{\junit}{JUnit}
\newcommand{\maven}{Maven}
\newcommand{\github}{Github}

\newcommand{\takeaway}[1]{
\begin{tcolorbox}[colback=white,sharp corners,boxrule=0.2mm]#1\end{tcolorbox}
}

\newcommand{\rev}[1]{\textcolor{black}{#1}}

\newcommand{\ie}{i.e.}

\newcommand{\TODO}[1]{\textcolor{red}{#1}\GenericWarning{}{LaTeX Warning: TODO: #1}}\newcommand\todo\TODO

\title{A Comprehensive Study of Pseudo-tested Methods}

\author{Oscar Luis Vera-Pérez \and Benjamin Danglot \and Martin Monperrus \and Benoit Baudry }
\institute{
	O. Vera-Pérez \at
	Inria Rennes - Bretagne Atlantique \\
	Campus de Beaulieu, 263 Avenue Général Leclerc \\
	35042 Rennes - France \\
	\email{oscar.vera-perez@inria.fr}
	\and
	B. Danglot \at
	Inria Lille - Nord Europe
	Parc scientifique de la Haute Borne \\ 
	40, avenue Halley - Bât A - Park Plaza \\
	59650 Villeneuve d'Ascq - France\\
	\email{danglot@inria.fr}
	\and
	M. Monperrus \at
	KTH Royal Institute of Technology in Stockholm\\
	Brinellvägen 8\\
	114 28 Stockholm - Sweden\\
	\email{martin.monperrus@csc.kth.se}
	\and
	B. Baudry \at
	KTH Royal Institute of Technology in Stockholm\\
	Brinellvägen 8\\
	114 28 Stockholm - Sweden\\
	\email{baudry@kth.se}
}

\begin{document}

\maketitle

% !TEX encoding = UTF-8
% !TEX spellcheck = en_US
% !TEX root = main.tex
% !TEX ts-program = latexmk

\begin{abstract}
Pseudo-tested methods are defined as follows: they are covered by the test suite, yet no test case fails when the method body is removed, i.e., when all the effects of this method are suppressed. 
This intriguing concept was coined in 2016, by Niedermayr and colleagues, who showed that such methods are systematically present, even in well-tested projects with high statement coverage. 

This work presents a novel analysis of \pseudotested{} methods. 
First, we run a replication of Niedermayr's study with 28K+ methods, enhancing its external validity thanks to the use of new tools and new study subjects.
Second, we perform a systematic characterization of these methods, both quantitatively and qualitatively with an extensive manual analysis of 101 \pseudotested{} methods.

The first part of the study confirms Niedermayr's results: \pseudotested{} methods exist in all our subjects. 
Our in-depth characterization of \pseudotested{} methods leads to two key insights: \pseudotested{} methods are significantly less tested than the other methods; yet, for most of them, the developers would not pay the testing price to fix this situation.
This calls for future work on targeted test generation to specify those pseudo-tested methods without spending developer time. 
\end{abstract}

\section{Introduction}

Niedermayr and colleagues \cite{niedermayr_will_2016} recently introduced the concept of \emph{\pseudotested{} methods}. 
These methods are covered by the test suite, but no test case fails even if all behaviors of the method are removed at once, i.e. when the body is completely stripped off.
This work is novel and intriguing:
such \pseudotested{} methods are present in all projects, even those with test suites that have high coverage ratio. 

If those results hold, it calls for more research on this topic.
This is the motivation of this paper: first, we challenge the external validity of Niedermayr et al.'s experiment with new study subjects, second we perform an in-depth qualitative empirical study of \pseudotested{} methods. 
In particular, we want to determine if \pseudotested{} methods are indicators of badly tested code. While this seems to be intuitively true, we aim at quantifying this phenomenon.
Second, we want to know whether \pseudotested{} methods are relevant indicators for developers who wish to improve their test suite. 
In fact, these methods may encapsulate behaviors that are poorly specified by the test suite, but are not relevant functionalities for the project.

To investigate \pseudotested{} methods, 
we perform an empirical study based on the analysis of 21 open source Java projects. In total, we analyze 28K+ methods in these projects.
We articulate our study around three parts.

In the first part we characterize our study subjects by looking at the number and proportion of \pseudotested{} methods. 
This also acts as a conceptual replication \cite{shull_role_2008} of Niedermayr's study. 
Our results mitigate two threats to the validity of Niedermayr's results: our methodology mitigates internal threats, by using another tool to detect \pseudotested{} methods, and our experiment mitigates external threats by augmenting Niedermayr's set of study objects with 17 additional open source project.

In the second part, we quantify the difference between \pseudotested{} methods and the other covered methods. We compare both sets of methods with respect to the fault detection ratio of the test suite and prove that \pseudotested{} methods are significantly worse tested with respect to this criterion.

In the third part, we aim at collecting a qualitative feedback from the developers. First, we manually found 	a set of pseudo-tested methods that reveal specific issues in the test suites of 7 projects. Then, we submitted pull requests and sent emails to the developers, asking their feedback about the relevance of these test issues. All pull requests have been merged to improve the test suite. Second, we met with the developers of 3 projects and inspected together a sample of 101 \pseudotested{} methods.  We found that developers consider only 30 of them worth  spending time improving the test suite.

To summarize, the contributions of this paper are as follow:
\begin{itemize}
	\item a conceptual replication of Niedermayr's initial study on \pseudotested{} methods. Our replication confirms the generalized presence of such methods, and improves the external validity of this empirical observation.
	\item a quantitative analysis of \pseudotested{} methods, which measures how different they are compared to other covered, not \pseudotested{} methods.
	\item a qualitative manual analysis of 101 \pseudotested{} methods, involving developers, that reveals that less than 30\% of these methods are clearly worth the additional testing effort.
	\item open science, with a complete replication package available at: 
	\url{https://github.com/STAMP-project/descartes-experiments/}.
\end{itemize}   

The rest of this paper is organized as follows.
Section \ref{sec:pseudo} defines the key concepts that form the core of this empirical study. 
Section \ref{sec:expe} introduces the research questions that we investigate in this paper, as well as the set of study subjects and the metrics that we collect.
In Section \ref{sec:results}, we present and discuss the observations for each research question. 
In Section \ref{sec:threats}, we discuss the threats to the validity of this study and Section \ref{sec:related} discusses related works.

\section{Pseudo-tested Methods}
\label{sec:pseudo}

In this section, we first define the key concepts that we investigate in this study. 
Our definitions are founded on the  presentation given by \theoriginalauthors{}. Then, we describe the procedure that we elaborate, in order to automatically detect \pseudotested{} methods. 

\subsection{Definitions}

Here we define the main interactions between a program $P$ and its test suite $TS$, which form the basis of all dynamic analyses presented in this paper. From now on, we consider a program $P$ to be a set of methods.

\begin{definition}
	\label{def:test-case}
	A test case $t$ is a method that initializes the program in a specific state, triggers specific behaviors and specifies the expected effects for these behaviors through assertions. \rev{$TS$ is the set of all test cases $t$ written for a program $P$}.
\end{definition}

\begin{definition}
	\label{def:covered-method}
	A method  $m \in P$ is said to be covered if there exists at least one test case $t \in TS$ that triggers the execution of at least one path of the body of $m$.
\end{definition}

\autoref{list:test-case} illustrates an example of a test case according to \autoref{def:test-case}. This test case covers the \texttt{getInstance}, \texttt{getCharRanges}, and \texttt{toString} methods of the \texttt{CharSet} class. It initializes the \texttt{CharSet} object by calling \texttt{getInstance} and setting an initial state. Then, the \texttt{getCharRanges} method is invoked to obtain the character ranges composing the set specification. The three assertions in lines \ref{assert1} - \ref{assert3} specify the expected effects of the method invocations. Since at least one path is executed we say that \texttt{getInstance} is covered by \texttt{testConstructor} as stated in \autoref{def:covered-method}. (Yet, not all the paths are covered, such as the instructions in lines \ref{null} and \ref{common}.)

As per \autoref{def:test-case}, in a Java program, a set of a \junit{} test classes is a particular case of a test suite. 

\begin{lstlisting}[caption=A test case covering two methods of the class CharSet taken from \texttt{commons-lang}, captionpos=b, label=list:test-case]
class CharSetTest {
    ...
    @Test
    public void testConstructor() {
        CharSet set = CharSet.getInstance("a");
        CharRange[] array = set.getCharRanges();
        assertEquals("[a]", set.toString()); %*\label{assert1}*)
        assertEquals(1, array.length);
        assertEquals("a", array[0].toString()); %*\label{assert3}*)
    }
    ...
}

class CharSet {
    ...
    public static CharSet getInstance(final String... setStrs) {
        if (setStrs == null) {
            return null;  %*\label{null}*)
        }
        if (setStrs.length == 1) {
            final CharSet common = COMMON.get(setStrs[0]);
            if (common != null) {
                return common;  %*\label{common}*)
            }
        }
        return new CharSet(setStrs);
    }
    ...
    public CharRange[] getCharRanges() { ... }
}
\end{lstlisting}

\rev{}

Let $m$ be a method; $S=\cup_{m \in P}{\mathit{effects}(m)}$ the set of effects of all methods in $P$; $\mathit{effects}(m)$ a function $\mathit{effects}:P \rightarrow S$ that returns all the effects of a method $m$; $detect$, a predicate $TS \times S \rightarrow \{\top, \bot\}$ that determines if an effect is detected by $TS$. \rev{Here, we consider the following possible effects that a method can produce: change the state of the object on which it is called, change the state of other objects (by calling other methods), return a value as a result of its computation.}

\begin{definition}
\label{def:pseudo}
	A method is said to be \pseudotested{} with respect to a test suite, if the test suite covers the method and does not assess any of its effects:
	\[ \forall s \in \mathit{effects}(m), \nexists t \in TS : detect(t, s) \]
\end{definition}

\begin{definition}
	\label{def:required}
	A method is said to be required if the test suite covers the method and assesses at least one of its effects:
	\[ \exists s \in \mathit{effects}(m), \exists t \in TS : detect(t, s) \]
\end{definition}

\begin{lstlisting}[caption=Example of a \pseudotested{} method, captionpos=b, label=list:pt-example]
class VList {
	private List elements;
	private int version;
	public void add(Object item) {
		elements.add(item);
		incrementVersion();
	}

	private void incrementVersion() { %*\label{line-ptm}*)
		version++;
	}

	public int size() {
		return elements.size();
	}
}

class VListTest {
	@Test
	public void testAdd() { %*\label{line-tc}*)
		VList l = new VList();
		l.add(1);
		assertEquals(l.size(), 1);
	}
}
\end{lstlisting}

\rev{\autoref{list:pt-example} shows an example of a \pseudotested{} method. \texttt{incrementVersion}, declared in line \ref{line-ptm}, is \pseudotested{}. The test case in line \ref{line-tc} triggers the execution of the method but does not assess its effect: the modification of the \texttt{version} field. In the absence of other test cases, the body of this method could be removed and the test suite will not notice the change. This particular example also shows that \pseudotested{} methods may pose a testing challenge derived from testability issues in the code.}

One can note that \theoriginalauthors{} call required  methods, ``tested methods''. We do not keep this terminology here, for two key reasons: (i) \rev{these covered methods may include behaviors that are not specified by the test suite, hence not all the effects haven been tested}; (ii) \rev{meanwhile, by contrast to \pseudotested{} methods, the correct execution of these methods is \emph{required} to make the whole test suite pass correctly since at least one effect of the method is assessed}.

\subsection{Tool for Finding Pseudo-tested Methods}
\label{sec:description-tool}

A ``pseudo-tested'' method, as defined previously, is an idealized concept.
In this section, we describe an algorithm that implements a practical way of collecting a set of \pseudotested{} methods in a program $P$, in the context of the test suite $TS$, based on the original proposal of \theoriginalauthors. 
It relies on the idea of ``extreme code transformations'', which consists in completely stripping out the body of a method.

Algorithm \ref{alg:detection} starts by analyzing all methods of $P$ that are covered by the test suite \rev{and fulfill a predefined selection criterion (predicate $\mathrm{ofInterest}$ in \autoref{of-interest}). This critetion is based on the structure of the method and aims at reducing the number of false positives detected by the procedure. It eliminates uninteresing methods such as trivial setter and getters or empty void methods. More insight on this will be given in Section \ref{sec:metrics}.}   
If the method returns a value, the body of the method is stripped out and we generate a few variants that simply return predefined values (\autoref{if-return}). These values depend on the return type, and are shown in \autoref{tab:desops}, 
\footnote{Compared to \theoriginalauthors, we add two new transformations, one to return $null$ and another to return an empty array. These additions allow to expand the scope of methods to be analyzed.}
If the method is void, we strip the body without further action (\autoref{no-return}).
Once we have a set of variants, we run the test suite on each of them, if no test case fails on any of the variants of a given method, we consider the method as \pseudotested{} (\autoref{is-pseudo}). \rev{One can notice in \autoref{alg-porig} that all extreme transformations are applied to the original program and are analyzed separately.}

\begin{algorithm}[t]
	\DontPrintSemicolon
	\KwData{$P$, $\mathit{TS}$}
	\KwResult{$pseudo$: \{\pseudotested{} methods in $P$\}}
	\ForEach{$m \in P | \mathrm{covered}(m,\mathit{TS}) \land \mathrm{ofInterest}(m)$}
	{\label{of-interest}	
		$variants$ : \{extreme variants of $m$\}\;
		\If{$\mathrm{returnsValue}(m)$} 
		{\label{if-return}
			stripBody($m$)\;
			checkReturnType($m$)\; 
			$variants \leftarrow \mathrm{fixReturnValues}(m)$\;
		}
		\Else{\label{no-return}
			$variants \leftarrow \mathrm{stripBody}(m)$\;
		}
		$failure \leftarrow false$\;
		\ForEach{$v \in variants$}
		{
			$P' \leftarrow \mathrm{replace}(m,v,P)$\;\label{alg-porig}
			$failure \leftarrow failure \lor \mathrm{run}(\mathit{TS},P')$\;	
		}
		\If{$\lnot failure$}{\label{is-pseudo}
			$pseudo \leftarrow pseudo \cup m$ 
		}
	}
	\KwRet pseudo
	\caption{Procedure to detect \pseudotested{} methods}
	\label{alg:detection}
\end{algorithm}

\begin{table}
	\caption{Extreme transformations used depending on return type.}
	\label{tab:desops}
	\centering
	\begin{tabular}{ll}
		\toprule
		Method type         & Values used \\
		\midrule
		void                & -           \\
		Reference types     & null        \\
		boolean             & true,false  \\
		byte,short,int,long & 0,1         \\
		float,double        & 0.0,0.1     \\
		char                & ` ', `A'   \\
		String              & ``'', ``A''\\
		T[]                 & new T[]\{\} \\
		\bottomrule
	\end{tabular}
\end{table}

To conduct our study, we have implemented Algorithm \ref{alg:detection} in an open source tool called Descartes\footnote{\url{https://github.com/STAMP-project/pitest-descartes}}. 
The tool can detect \pseudotested{} methods in Java programs tested with a JUnit test suite.
Descartes is developed as an extension of \pit{} \cite{coles_pit_2016}. \rev{\pit{} is a state-of-the-art mutation testing tool for \java{} projects that works with all major build systems such as Ant Gradle and Maven. This mutation tool is under active maintenance and development. It provides several features for test running and selection and can be extended via plugins. Descartes leverages the maturity of \pit{} and handles the discovery of points where extreme transformations can be applied and the creation of the new program variants \cite{veraperez2018descartes}.}
\rev{As a byproduct of the analysis and the features provided by \pit{}, Descartes is able to report the covered methods according to \autoref{def:covered-method}.}
Being open-source, we hope that Descartes  will be used by future research on the topic of \pseudotested{} methods .

\section{Experimental Protocol}
\label{sec:expe}

Pseudo-tested methods are intriguing. They are covered by the test suite, their body can be drastically altered and yet the test suite does not notice the transformations. We design an experimental protocol to explore the nature of those \pseudotested{} methods.

\subsection{Research Questions}

Our work is organized around the following research questions:
\begin{itemize}
	\item \textbf{RQ1} \textit{How frequent are \pseudotested{} methods?} This first question aims at characterizing the prevalence of \pseudotested{} methods. It is a conceptual replication of the work by \theoriginalauthors, with a larger set of study objects and a different tool support for the detection of \pseudotested{} methods.
	\item \textbf{RQ2} \textit{Are \pseudotested{} methods the weakest points in the program, with respect to the test suite?} This second question aims at determining to what extent \pseudotested{} methods actually capture areas in the code that are less tested than other parts. Here we use the mutation score as a standard test quality assessment metric. We compare the method-level mutation score of \pseudotested{} methods against that of other covered and required  methods (that are not \pseudotested). \rev{Here by method-level mutation score we mean the mutation score computed for each method,  considering only the mutants created inside each particular method.}
	\item \textbf{RQ3} \textit{Are \pseudotested{} methods helpful for developers to improve the quality of the test suite?} In this question we manually identify eight issues in the test suites of seven projects. We communicate these issues to the development teams through pull requests or email and collect their feedback.
	\item \textbf{RQ4} \textit{Which \pseudotested{} methods do developers consider worth an additional testing action?} Following our exchange with the developers, we expand the qualitative analysis to a sample of 101 \pseudotested{} methods distributed across three of our study subjects. We consulted developers to characterize the \pseudotested{} methods that are worth an additional testing action and the ones that are not worth it.
\end{itemize}

\subsection{Study Subjects}

\begin{table}[t]
	\caption{Projects used as study subjects. $\blacktriangle$: Projects taken from the work of \theoriginalauthors. $\blacktriangledown$: Projects taken from our  industry partners. $\blacklozenge$: Other projects used in the software testing literature. $\blacksquare$: Project from Github with more than 12,000 commits. $\bigstar$: Projects with more than a million LOC}
	\label{tab:projects}
	\begin{adjustbox}{max width=\textwidth,center=\textwidth}
	\begin{tabular}{lln{7}{0}n{5}{0}n{5}{0}}
		\toprule  
		Project & ID & {App LOC} & {Test LOC} & {Tests} \\
		\midrule
		$\blacktriangledown$ AuthZForce PDP Core          & \texttt{authzforce}          &   12596 &   3463 &   634 \\
		$\bigstar$           Amazon Web Services SDK      & \texttt{aws-sdk-java}       & 1676098 &  24115 &  1291 \\
		$\blacklozenge$      Apache Commons CLI           & \texttt{commons-cli}         &    2764 &   4241 &   460 \\
		$\blacklozenge$      Apache Commons Codec         & \texttt{commons-codec}       &    6485 &  10782 &   663 \\
		$\blacktriangle$     Apache Commons Collections   & \texttt{commons-collections} &   23713 &  27919 & 13677 \\
		$\blacklozenge$      Apache Commons IO            & \texttt{commons-io}          &    8839 &  15495 &   963 \\
		$\blacktriangle$     Apache Commons Lang          & \texttt{commons-lang}        &   23496 &  37237 &  2358 \\
		$\blacklozenge$      Apache Flink                 & \texttt{flink-core}          &   46390 &  30049 &  2341 \\
		$\blacklozenge$      Google Gson                  & \texttt{gson}                &    7184 &  12884 &   951 \\
		$\blacklozenge$      Jaxen XPath Engine           & \texttt{jaxen}               &   12467 &   8579 &   716 \\
		$\blacktriangle$     JFreeChart                   & \texttt{jfreechart}          &   94478 &  39875 &  2138 \\
		$\blacklozenge$      Java Git                     & \texttt{jgit}                &   75893 &  52981 &  2760 \\
		$\blacklozenge$      Joda-Time                    & \texttt{joda-time}           &   28724 &  55311 &  4207 \\
		$\blacklozenge$      JOpt Simple                  & \texttt{jopt-simple}         &    2386 &   6828 &   817 \\
		$\blacklozenge$      jsoup                        & \texttt{jsoup}               &   11528 &   6311 &   561 \\
		$\blacktriangledown$ SAT4J Core                   & \texttt{sat4j-core}          &   18310 &   8091 &   710 \\
		$\blacklozenge$      Apache PdfBox                & \texttt{pdfbox}              &  121121 &  15978 &  1519 \\
		$\blacksquare$       SCIFIO                       & \texttt{scifio}              &   49005 &   6342 &  1021 \\
		$\blacktriangledown$ Spoon                        & \texttt{spoon}               &   48363 &  32833 &  1371 \\
		$\blacktriangle$     Urban Airship Client Library & \texttt{urbanairship}        &   25260 &  15625 &   701 \\
		$\blacktriangledown$ XWiki Rendering Engine       & \texttt{xwiki-rendering}     &   37571 &   9276 &  2247 \\
		\midrule
		Total                                             &                     & 2332671 & 424215 & 42106 \\
		\bottomrule
	\end{tabular}
\end{adjustbox}
\end{table}

We selected 21 open source projects in a systematic manner to conduct our experiments. We considered active projects written in \java{}, that use \maven{} as main build system, \junit{} as the main testing framework and their code is available in a version control hosting service, mostly \github{}. 

A project is selected if it meets one of the following conditions:
1) they are present in the experiment of \theoriginalauthors{} (4 projects),
2) they are maintained by industry partners from whom we can get qualitative feedback (4 projects),
3) they are regularly used in the software testing literature (11 projects),
4) they have a mature history with more than 12,000 commits (one project) or they have a code base surpassing one million lines of code (one project).

Table \ref{tab:projects} shows the entire list. The first two columns show the name of each project and the identifiers we use to distinguish them in the present study. The third and fourth columns of this table show the number of lines of code in the main code base and testing code respectively. Both numbers were obtained using \textit{cloc}\footnote{\url{https://github.com/AlDanial/cloc}}. The last column shows the number of test cases as reported by \maven{} when running \textit{mvn test}. 
For instance, Apache Commons IO (\texttt{commons-io}) has \numprint{8839} lines of application code and \numprint{3463} lines of test code spread over 634 test cases.
The smallest project, Apache Commons Cli (\texttt{commons-cli}), has \numprint{2764} lines, while the largest, Amazon Web Services (\texttt{aws-sdk-java}) is composed of 1.6 million lines of Java code.
The list shows that our inclusion criteria enable us to have a wide diversity in terms of project size. In many cases the test code is as large or larger than the application code base.

We have prepared a replication package at: 

\burl{https://github.com/STAMP-project/descartes-experiments/}.

\subsection{Metrics}
\label{sec:metrics}

In this section, we define the metrics we used to perform the quantitative analysis of \pseudotested{} methods.

\textbf{Number of methods (\#METH).} The total number of methods found in a project, after excluding  constructors and static initializers. We make this choice because these methods cannot be targeted by our extreme transformations.

Certain types of methods are not generally targeted by developers in unit tests, we exclude them to reduce the number of methods that developers may consider as false positives. In our analysis, we do not consider:
\begin{itemize}
	\item methods that are not covered by the test suite. We ignore these methods, since, by definition, \pseudotested{} methods are covered.
	\item \texttt{hashCode} methods, as suggested by \theoriginalauthors, since this type of transformation would still convey with the hash code protocol
	\item methods with the structure of a simple getter or setter (\ie{} methods that only return the value of an instance field or assign a given value to an instance field), methods marked as deprecated \rev{(\ie{} methods explicitly marked with the \texttt(@Deprecated) annotation or declared in a class marked with that same annotation)}, empty void methods, methods returning a literal constant and compiler generated methods such as synthetic methods or methods generated to support $enum$ types
\end{itemize}

\textbf{Number of methods under analysis (\#MUA).} Given a program $P$, that includes \#METH methods, the number of methods under analysis \#MUA is obtained after excluding the methods described above. 

\textbf{Ratio of covered methods (C\_RATE).} A program $P$ can be seen as a set of methods. When running a test suite $TS$ on $P$ a subset of methods $COV \subset P$ are covered by the test suite. The ratio of covered methods is defined as C\_RATE$=\frac{|COV|}{|P|}$. \rev{In practice, $COV$ is computed by our tool, as explained in Section \ref{sec:description-tool}.} 

\textbf{Ratio of \pseudotested{} methods (PS\_RATE).} For all methods under analysis, we run the procedure described in Algorithm \ref{alg:detection}, to get the subset of \pseudotested{} methods, noted as \#PSEUDO methods.
The ratio of \pseudotested{} methods of a program is defined as PS\_RATE=$\frac{\mathrm{\#PSEUDO}}{\mathrm{\#MUA}}$.

\rev{PS\_RATE is used in RQ1 to determine the presence of \pseudotested{} methods in our study subjects. We also use the Pearson coefficient to check if we can state a correlation between PS\_RATE and C\_RATE.}

\textbf{Mutation score}. Given a program $P$, its test suite $TS$ and a set of mutation operators $op$, a mutation tool  generates a set $M$ of mutants for $P$. 
The mutation score of the test suite $TS$ over $M$ is defined as the ratio of detected mutants included in $M$. That is:
\begin{equation}
	score(M) = \frac{|\mu: \mu \in M \land detected(\mu)|}{|M|}
\end{equation}	
where $detected(\mu)$ means that \rev{at least one test case in $TS$ fails when the suite is executed against $\mu$}.

\textbf{Mutation score for \pseudotested{} methods (MS\_pseudo)}: the score computed with the subset of mutants generated by applying the mutation operators only on \pseudotested{} methods.
	
\textbf{Mutation score for required methods (MS\_req)}: the score computed with the subset of mutants generated by applying the mutation operators only on required methods.

\rev{These three mutation score metrics are used in RQ2 to quantitatively determine if \pseudotested{} methods are the worst tested methods in the code base. We perform a Wilcoxon statistical test to compare the values obtained for MS\_pseudo and MS\_req on our study subjects.\footnote{The computation of the Pearson coefficient and the Wilcoxon test were performed using the features of the \texttt{R} language.}}

\section{Experimental Results}
\label{sec:results}

The following sections present in depth our experimental results.

\subsection{RQ1: How frequent are \pseudotested{} methods?}
\label{sec:rq1}

We analyzed each study subject following the procedure described in Section \ref{sec:description-tool}. The results are summarized in Table \ref{tab:testedclass}. The second column shows the total number of methods excluding constructors. The third, lists the methods covered by the test suite. The following column shows the ratio of covered methods. The \textquote{\#MUA} column shows the number of methods under analysis, per the criteria described in Section \ref{sec:metrics}. The last two columns give the number of \pseudotested{} methods (\#PSEUDO) and their ratio to the methods under analysis (PS\_RATE).

For instance, in \texttt{authzforce}, 325 methods are covered by the test suite, of which 291 are relevant for the analysis. In total, we identify 13 \pseudotested{} methods representing 4\% of the methods under analysis.

We discover \pseudotested{} methods in all our study objects, even for those with high coverage. This corroborates the observations made by \theoriginalauthors.
The number of observed \pseudotested{} methods ranges from 2 methods, in \texttt{commons-cli}, to 473 methods in \texttt{jfreechart}.
The PS\_RATE varies from 1\% to 46\%. In 14 cases its value remains below 7\% of all analyzed methods. This means that, compared to the total number of methods in a project, the amount of \pseudotested{} methods can be managed by the developers in order to guide the improvement of test suites.

\subsubsection{Analysis of outliers}
Some projects have specific features that may affect the PS\_RATE value in a distinctive way. In this section we discuss those cases.

\texttt{authzforce}, uses almost exclusively parameterized tests. The ratio of covered methods is low, but these methods have been tested exhaustively and therefore they have a lower PS\_RATE compared to other projects with similar coverage.

The application domain can have an impact on the design of test suites. For example, \texttt{scifio} is a framework that provides input/output functionalities for image formats that are typically used in scientific research. This project shows the highest PS\_RATE in our study. When we look into the details, \label{sec:scifio-problem} we find that 62 out of their 72 \pseudotested{} methods belong to the same class and deal with the insertion of metadata values in DICOM images, a format widely used in medical imaging. Not all the metadata values are always required and the test cases covering these methods do not check their presence. \texttt{pdfbox} is another interesting example in this sense. It is a library designed for the creation and manipulation of PDF files. Some of their functionalities can only be checked by visual means which increases the level of difficulty to specify an automated and fine-grained oracle. Consequently, this project has a high PS\_RATE.

At the other end of the  PS\_RATE spectrum, we find \texttt{commons-cli} and \texttt{jopt-simple}. These are small projects, similar in purpose and \rev{both have comprehensive test suites that reach 97\% and 98\% of line coverage respectively (as measured by \textit{cobertura}\footnote{\url{http://cobertura.github.io/cobertura/}}).} Only two \pseudotested{} methods were found for each one of them. Three of those four methods create and return an exception message. The remaining method is a \texttt{toString} implementation.

\subsubsection{Relationship between \pseudotested{} methods and coverage}

We observe that the projects with lowest method coverage show higher ratios of \pseudotested{} methods. The Pearson coefficient between the coverage ratio and the ratio of \pseudotested{} methods is -0.67 and $p < 0.01$ which indicates a moderate negative relationship.

\rev{This confirms our intuition that \pseudotested{} methods are more frequent in projects that are poorly tested (high \pseudotested{} ratios and low coverage ratios).} However, the ratio of \pseudotested{} methods is more directly impacted by the way the methods are verified and not the ratio of methods covered. It is possible to achieve a low ratio of \pseudotested{} methods covering a small portion of the code. For example, \texttt{authzforce} and \texttt{xwiki-rendering} have comparable coverage ratios but the former has a lower ratio of \pseudotested{} methods. The correlation with the ratio is a consequence of the fact that, in general, well tested projects also have higher coverage ratios. 

\subsubsection{Comparison with the study by \theoriginalauthors{}}

Our study, on a new dataset, confirms the major finding of \theoriginalauthors's study:
\pseudotested{} methods exist in all projects, even the very well tested ones.
This first-ever replication improves the external validity of this finding.
We note in the original study by \theoriginalauthors{}, that the reported ratio was higher, ranging from 6\% to 53\%. The difference can be explained from the fact that 1)
we exclude deprecated methods and 
2) we consider two new other mutation operators. These two factors change the set of methods that have been targeted.

\begin{table}
    \caption{Number of methods in each project, number of methods under analysis and number of \pseudotested{} methods}
	\label{tab:testedclass}
	\begin{adjustbox}{max width=\textwidth,center=\textwidth}
    \begin{tabular}{ln{7}{0}n{4}{0}rn{4}{0}n{4}{0}r}
        \toprule
        Project & {\#Methods} & {\#Covered} & C\_RATE & {\#MUA} & {\#PSEUDO}  & PS\_RATE  \\
        \midrule
        \texttt{authzforce}          &    697 &  325 & 47\% &  291 &   13 &  4\% \\  
		\texttt{aws-sdk-java}        & 177449 & 2314 &  1\% & 1800 &  224 & 12\% \\
		\texttt{commons-cli}         &    237 &  181 & 76\% &  141 &    2 &  1\% \\  
		\texttt{commons-codec}       &    536 &  449 & 84\% &  426 &   12 &  3\% \\  
		\texttt{commons-collections} &   2729 & 1270 & 47\% & 1232 &   40 &  3\% \\  
		\texttt{commons-io}          &    875 &  664 & 76\% &  641 &   29 &  5\% \\  
		\texttt{commons-lang}        &   2421 & 1939 & 80\% & 1889 &   47 &  2\% \\  
		\texttt{flink-core}          &   4133 & 1886 & 46\% & 1814 &  100 &  6\% \\  
		\texttt{gson}                &    624 &  499 & 80\% &  477 &   10 &  2\% \\    
		\texttt{jaxen}               &    958 &  616 & 64\% &  569 &   11 &  2\% \\  
		\texttt{jfreechart}          &   7289 & 3639 & 50\% & 3496 &  476 & 14\% \\  
		\texttt{jgit}                &   6137 & 3702 & 60\% & 2539 &  296 & 12\% \\  
		\texttt{joda-time}           &   3374 & 2783 & 82\% & 2526 &   82 &  3\% \\  
		\texttt{jopt-simple}         &    298 &  265 & 89\% &  256 &    2 &  1\% \\  
		\texttt{jsoup}               &   1110 &  844 & 76\% &  751 &   28 &  4\% \\  
		\texttt{sat4j-core}          &   2218 &  613 & 28\% &  585 &  143 & 24\% \\  
		\texttt{pdfbox}              &   8164 & 2418 & 30\% & 2241 &  473 & 21\% \\  
		\texttt{scifio}              &   3269 &  895 & 27\% &  158 &   72 & 46\% \\  
		\texttt{spoon}               &   4470 & 2976 & 67\% & 2938 &  213 &  7\% \\  
		\texttt{urbanairship}        &   2933 & 2140 & 73\% & 1989 &   28 &  1\% \\  
		\texttt{xwiki-rendering}     &   5002 & 2232 & 45\% & 2049 &  239 & 12\% \\
		\midrule 
		Total               & 234923 & 32650 & 14\% & 28808 & 2540 & 9\% \\
        \bottomrule
	\end{tabular}
\end{adjustbox}
\end{table}

\takeaway{We have made the first independent replication of \theoriginalauthors's study. Our replication confirms that all Java projects contain \pseudotested{} methods, even the very well tested ones. This improves the external validity of this empirical fact. The ratio of \pseudotested{} methods with respect to analyzed methods ranged from 1\% to 46\% in our dataset.}

\subsection{RQ2: Are \pseudotested{} methods the weakest points in the program, with respect to the test suite?}
\label{sec:rq2}

By definition, test suites fail to assess the presence of any effect in \pseudotested{} methods. As such, these methods can be considered as very badly tested, even though they are covered by the test suite. To further confirm this fact we assess the test quality of these methods with a traditional test adequacy criterion: mutation testing \cite{demillo1978hints}. To do so, we measure the chance for a mutant planted in a \pseudotested{} method to be detected (killed).

For each of our study subjects, we run a mutation analysis based on \pit{}, a state of the art mutation tool for Java. We configure \pit{} with its standard set of mutation operators. \pit{} is capable of listing: the comprehensive set of mutants, the method in which they have been inserted and whether they have been detected (killed) by the test suite. 
We  extract the set of mutants that have been placed in the body of the \pseudotested{} methods to compute the mutation score on those methods (MS\_pseudo) as well as the mutation score of required methods (MS\_req). 

\begin{figure}
	\caption{Mutation score for mutants placed inside \pseudotested{} and required methods.}
	\label{fig:scores}
	\includegraphics[scale=.6]{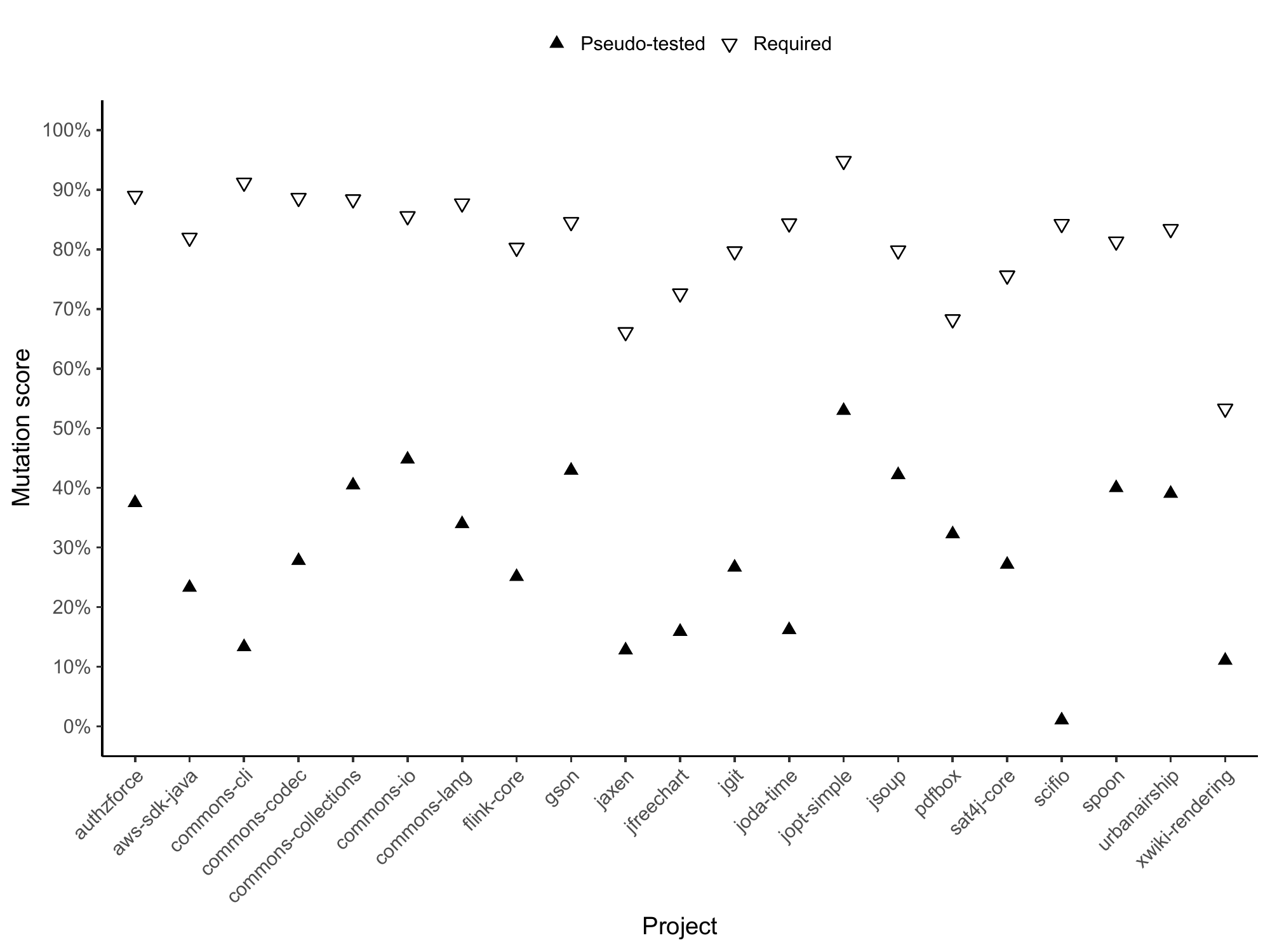}
\end{figure}

Figure \ref{fig:scores} shows the results of this experiment.
In all cases, the mutation score of \pseudotested{} methods is significantly lower than the score of normal required methods. This means that a mutant planted inside a \pseudotested{} method has more chances to survive than a mutant planted in required methods. The minimum gap is achieved in \texttt{pdfbox} with scores 32\% for \pseudotested{} methods and 68\% for others. For \texttt{scifio}, only 1\% of \pit{} mutants in \pseudotested{} methods can be killed (as opposed to 84\% in required methods).
To validate this graphical finding, we compare  MS\_pseudo and MS\_req.

\rev{In average the mutation score of required methods is 52\% above that of pseudo-tested methods.
With the Wilcoxon statistical test, this is a significant evidence of a difference with a p-value $p < 0.01$. The effect size is 1.5 which is considered as large per the standard guidelines in software engineering \cite{KampenesDHS07}.}

\subsubsection{Analysis of interesting examples}

It calls the attention, that, in no case, MS\_pseudo was 0\%. So, even when extreme transformations are not spotted by the test suite, some mutants inside these methods can be detected. We now explain this case.

Listing \ref{list:ps-with-detected} shows a simplified extract of a \pseudotested{} method we have found in \texttt{auzthforce} and where some traditional mutants were detected. The \texttt{checkNumberOfArgs} method is covered by six test cases and was found to be \pseudotested{}. In all test cases, the value of \texttt{numInputs} is greater than two, hence the condition on line \ref{ln:condition} was always false and the exception was never thrown. \pit{} created five mutants in the body of this method and two of them were detected. Those mutants replaced the condition by $true$ and $<$ by $<=$ respectively. With this, the condition is always $false$, the exception is thrown and the mutants are detected. It means that those mutants are trivially detected with an exception, not by an assertion. This is the major reason for which the mutation score of \pseudotested{} methods can be higher than 0.

This explanation holds for \texttt{jopt-simple} which achieves a seemingly high  53\% MS\_pseudo. A total of 17 mutants are generated in the two \pseudotested{} methods of the project. Nine of these mutants are killed by the test suite. From these nine, six replaced internal method calls by a default value and the other three replaced a constant by a different value. All nine mutations made the program crash with an exception, and are thus trivially detected.

\begin{lstlisting}[caption=A pseudo-tested method where traditional mutants were detected, language=java, captionpos=b, label=list:ps-with-detected]
class AnyOfAny {
  protected void checkNumberOfArgs(int numInputs) {
    if(numInputs < 2) %*\label{ln:condition}*)
      throw new IllegalArgumentException();
  }

  public void evaluate(... args) {
    checkNumberOfArgs(args.size())
    ...
  }
}
\end{lstlisting}

\texttt{scifio} has the lowest MS\_pseudo. \pit{} generated \numprint{7598} mutants in the 62 methods dealing with metadata and mentioned in Section \ref{sec:scifio-problem}. The mutants modify the metadata values to be placed in the image and, as discussed earlier, those values are not specified in any oracle of the test suite. Hence, none of these mutants are detected.

\subsubsection{Distribution of method-level mutation score}

\begin{figure}
	\caption{\pit{} method-level mutation score distribution by project and method category}
	\label{fig:distributions}
	\centering
	\includegraphics[scale=.6]{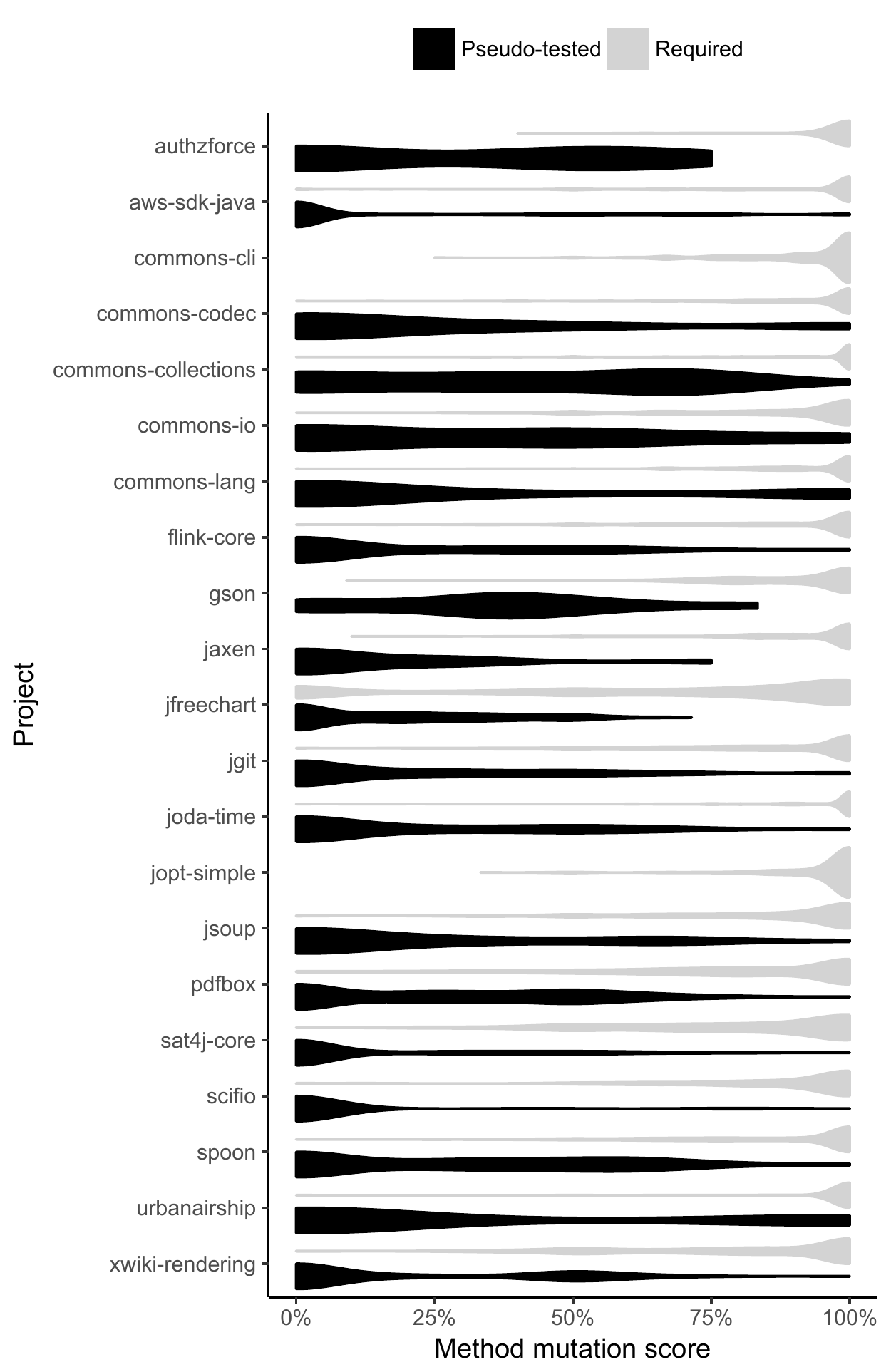}
\end{figure}

To further explore the difference between  MS\_pseudo and MS\_req, we compute the distribution of the method-level mutation score. That is, we compute a mutation score for each method in a project.
The final distribution for each project is shown in Figure \ref{fig:distributions}. Each row displays two violin plots for a specific project.\footnote{The violin plot for \pseudotested{} methods of \texttt{commons-cli} and \texttt{jopt-simple} are not displayed, as they have too few methods in this category.} Each curve in the plot represents the distribution of mutation scores computed per method. The thicker areas on each curve represent scores achieved by more methods. 

The main finding is that the distributions for the required methods are skewed to the right (high mutation score), while the scores for \pseudotested{} methods tend to be left skewed.
This is a clear trend, which confirms the results of \autoref{fig:scores}.
It is also the case that most distributions cover a wide range of values. While most \pseudotested{} methods have low scores, there are cases for which the mutation score could reach high values, due to trivial exception-raising mutants. 

In these plots the already discussed phenomenon for the methods of \texttt{scifio} also becomes visible. Those methods have each a considerable number of mutants and none of them were detected, therefore the \texttt{scifio} distribution for \pseudotested{} methods is remarkably left skewed.

We observe that 63 \pseudotested{} methods across all projects have a 100\% mutation score. Among those 63, 34 have only one or two trivial mutants.
\rev{As an extreme case, in the \texttt{jsoup} project we find that the method \texttt{load} of the \texttt{Entities} class\footnote{\url{https://github.com/jhy/jsoup/blob/35e80a779b7908ddcd41a6a7df5f21b30bf999d2/src/main/java/org/jsoup/nodes/Entities.java\#L295}} is \pseudotested{} and \pit{} generates 69 mutants that are all killed}. All mutants make the program crash with an exception, yet the body of the method can be removed and the absence of effects is unnoticed by the test suite. This suggests that the extreme transformations performed to find \pseudotested{} methods are less susceptible to be trivially detected.

\takeaway{The hypothesis that \pseudotested{} methods expose weakly tested regions of code is confirmed by mutation analysis. For all the 21 considered projects, the mutation score of \pseudotested{} methods is significantly lower than the score of required methods, a finding confirmed by a very low p-value lower than 0.01 and a very high effect size of 1.5.}

\subsection{RQ3: Are \pseudotested{} methods relevant for developers to improve the quality of the test suite?}
\label{sec:rq3}

To answer this question, we manually analyze the $void$ and $boolean$ \pseudotested{} methods which are accessible from an existing test class. $void$ and $boolean$ methods have only one or two possible extreme transformations, thus are easier to explain to developers. We identify eight testing issues revealed by these \pseudotested{} methods: two cases of a miss-placed oracle, two cases of missing oracle, three cases of a weak oracle and one case of a missing test input. These issues have been found in seven of our study subjects.

For each testing issue we prepare a pull request that fixes the issue, or we send the information by email. Our objective is to collect qualitative feedback from the development teams about the relevance of the testing issues revealed by \pseudotested{} methods. 

We now summarize the discussion about each testing issue.

\subsubsection{Feedback from \texttt{aws-sdk-java}}

Per our selection criterion, we have spotted one \pseudotested{} method.
We made one pull request (PR)\footnote{\url{https://github.com/aws/aws-sdk-java/pull/1437}} to explicitly assess the effects of one \pseudotested{} method named \texttt{prepareSocket}. This method is covered by four test cases that follow the same pattern. A simplified extract is shown in Listing \ref{list:aws-case}. The test cases mock a socket abstraction that verifies if a given array matches the expected value. \texttt{prepareSocket} should call the \texttt{setEnabledProtocols} in the socket abstraction. When running an extreme transformation on this method, the assertion is never evaluated and the test cases pass silently. In the pull request we moved the assertion out of the \texttt{setEnabledProtocols} method, in order to have it verified after \texttt{prepareSocket}. Listing \ref{list:aws-fixture} shows a simplified version of the proposed code. With this modification, the method is not \pseudotested{} anymore. 
The developer agreed that the proposed change was an improvement and the pull request was merged into the code. This is an example of a miss-placed oracle and the value of \pseudotested{} methods.

\begin{lstlisting}[caption=A weak test case for method prepareSocket., captionpos=b, label=list:aws-case]
@Test
void typical() {
	SdkTLSSocketFactory f = ...;
	//prepareSocket was found to be pseudo-tested
	f.prepareSocket(new TestSSLSocket() {
	  ...
	  @Override
	  public void setEnabledProtocols(String[] protocols) {
		assertTrue(Arrays.equals(protocols, expected));
	  }
	...
   });
}
\end{lstlisting}
	
\begin{lstlisting}[caption={Proposed test improvement. The assertion was moved out of the socket implementation. Consequently, \texttt{prepareSocket} is no longer \pseudotested}, captionpos=b, label=list:aws-fixture]
@Test
void typical() {
	SdkTLSSocketFactory f = ...;
	SSLSocket s = new TestSSlSocket() {
	@Override
	public void setEnabledProtocols(String[] protocols) {
		capturedProtocols = protocols;
	}
	...
	};
	f.prepareSocket(s);
	//This way the test fails if no protocol was enabled
	assertArrayEquals(s.capturedProtocols, expected));
}
\end{lstlisting}

\subsubsection{Feedback from \texttt{commons-collections}}

In this project, certain methods implementing iterator operations are found to be \pseudotested{}. Specifically two implementations of the \texttt{add} method and four of the \texttt{remove} method are  \pseudotested{} in classes where these operations are not supported. Listing \ref{list:collections-case} shows one the methods and a simplified extract of the test case designed to assess their effects. If the \texttt{add} method is emptied, then the exception is never thrown and the test passes. We proposed a pull request \footnote{\url{https://github.com/apache/commons-collections/pull/36}} with the change shown in Listing \ref{list:collections-fixture}. The proposed change verifies that an exception has been thrown. As in the previous example, the issue is related to the placement of the assertion. 
The developer agreed to merge the proposed test improvement into the code. This is a second example of the value of \pseudotested{} methods. Being in a different project, and assessed by another developer, this increases our external validity.

\begin{lstlisting}[caption=Class containing the pseudo-tested method and the covering test class., captionpos=b, label=list:collections-case]
	class SingletonListIterator 
	  implements Iterator<Node> {
	  ...
	  void add() {
		//This method was found to be pseudo-tested
		throw new UnsupportedOperationException();
	  }
	  ...
	}
	
	class SingletonListIteratorTest {
	  ...
	  @Test
	  void testAdd() {
		SingletonListIterator it = ...;
		...
		try {
		  //If the method is emptied, then nothing happens
		  //and the test passes.
		  it.add(value);
		} catch(Exception ex) {}
	  ...
	}
\end{lstlisting}

\begin{lstlisting}[caption=Change proposed in the pull request to verify the unsupported operation., captionpos=b, label=list:collections-fixture]
...
try {
	it.add(value);
	fail(); //If this is executed, 
			//then the test case fails
} catch(Exception ex) {}
...
\end{lstlisting}

\subsubsection{Feedback from \texttt{commons-codec}}

For \texttt{commons-codec} we found that the $boolean$ method \texttt{isEncodeEqual} was \pseudotested{}. The method is covered by only one test case, shown in Listing \ref{list:codec-case}. As one can notice, the test case lacks the corresponding assertion. \rev{So, none of the extreme transformations applied to this method could cause the test to fail.}

\begin{lstlisting}[caption=Covering test case with no assertion., captionpos=b, label=list:codec-case]
public void testIsEncodeEquals() {         
    final String[][] data = {
		{"Meyer", "M\u00fcller"}, %*\label{wrong-input}*)
		{"Meyer", "Mayr"},
		...
		{"Miyagi", "Miyako"}
	};
    for (final String[] element : data) { 
            final boolean encodeEqual = 
		this.getStringEncoder().isEncodeEqual(element[1], element[0]);             
 }     
}
\end{lstlisting}

All the inputs in the test case should make the method return $true$. When we placed the corresponding assertion we found that the first input (in line \ref{wrong-input}) was wrong and we replaced it by a correct pair of values. We made a pull request\footnote{\url{https://github.com/apache/commons-codec/pull/13}} and the fixture was accepted by the developers and also slightly increased the code coverage by 0.2\%.

\subsubsection{Feedback from \texttt{commons-io}}

In \texttt{commons-io} we found several $void$ \texttt{write} methods of the \texttt{TeeOutputStream} to be \pseudotested{}. This class represents an output stream that should send the data being written to other two output streams. A reduced version of the test case covering these methods can be seen in Listing \ref{list:io-case}. Line \ref{wrong-verification} shows that the assertion checks that both output streams should contain the same data. If the \texttt{write} method is emptied, nothing is written to both streams but the assertion remains valid as both have the same content (both are empty). The test case should verify not only that those two streams have the same content but that they have the right value. In this sense, we say that this is an example of a weak oracle. We made a pull request\footnote{\url{https://github.com/apache/commons-io/pull/61}} with the changes exposed in Listing \ref{list:io-fixture}. The change adds a third output stream to be used as a reference value. The pull request was accepted and it slightly increased the code coverage by 0.07\%.

\begin{lstlisting}[caption=Test case verifying TeeOutputStream write methods., captionpos=b, label=list:io-case]
public void testTee() {    
  ByteArrayOutputStream baos1 = new ByteArrayOutputStream();
  ByteArrayOutputStream baos2 = new ByteArrayOutputStream();
  TeeOutputStream tos = new TeeOutputStream(baos1, baos2);    
  ...    
  tos.write(array);
  assertByteArrayEquals(baos1.toByteArray(), baos2.toByteArray()); %*\label{wrong-verification}*)
}
\end{lstlisting}

\begin{lstlisting}[caption=Change proposed to verify the result of the write methods, captionpos=b, label=list:io-fixture]
public void testTee() {    
  ByteArrayOutputStream baos1 = new ByteArrayOutputStream();
  ByteArrayOutputStream baos2 = new ByteArrayOutputStream();
  ByteArrayOutputStream expected = new ByteArrayOutputStream();    
  TeeOutputStream tos = new TeeOutputStream(baos1, baos2);    
  ...    
  tos.write(array);
  expected.write(array);
  assertByteArrayEquals(expected.toByteArray(), baos1.toByteArray());
  assertByteArrayEquals(expected.toByteArray(), baos2.toByteArray());
}
\end{lstlisting}

For three projects, \texttt{spoon}, \texttt{flink-core} and \texttt{sat4j-core}, we discuss the details of the testing issues\footnote{\url{https://github.com/STAMP-project/descartes-experiments/blob/6f8a9c7c111a1da5794622652eae5327d0571ef1/direct-communications.md}} directly with the developers via emails.  We systematically collected their feedback.

\subsubsection{Feedback from \texttt{spoon}}
We ask the project team about a public $void$ method, \rev{named \texttt{visitCtAssert}\footnote{\url{https://github.com/INRIA/spoon/blob/fd878bc71b73fc1da82356eaa6578f760c70f0de/src/main/java/spoon/reflect/visitor/DefaultJavaPrettyPrinter.java\#L479}}}, and covered indirectly by only one test case. This method was part of a visitor pattern implementation, which is common inside this project. This particular method handles $assert$ \java{} expressions in an Abstract Syntax Tree. The test case does not assess the effects of the method. The developers expressed that this method should be verified by adding a stronger verification or a new test case. They were interested in our findings. They took no immediate action but opened a general issue \footnote{\url{https://github.com/INRIA/spoon/issues/1818}}.

\subsubsection{Feedback from \texttt{flink-core}} 

This team was contacted to discuss about a public $void$ method, \rev{named \texttt{configure}\footnote{\url{https://github.com/apache/flink/blob/740f711c4ec9c4b7cdefd01c9f64857c345a68a1/flink-core/src/main/java/org/apache/flink/api/common/io/BinaryInputFormat.java\#L86}}}, which is directly called by a single test case. This particular method loads a given configuration and prevents a field value from being overwritten. Listing \ref{list:flink-case} shows a simplified extract of the code. The body of the method could be removed and the test passes as the assertion only involves the initial value of the field. The developers explained that the test case was designed precisely to verify that the field is not changed after the method invocation. They expressed that more tests could probably make the scenario more complete. In our view, the test case should assert both facts: the configuration being loaded and the value not being changed. The current oracle expresses a weaker condition as the former verification is not done. If the body is erased, the configuration is never loaded and the value, of course, is never changed. The developers did not take any further action.

\begin{lstlisting}[caption=Pseudo-tested method in \texttt{flink-core}, captionpos=b, label=list:flink-case]
public void configure(Configuration parameters) {
	super.configure(parameters);
	if(this.blockSize == NATIVE_BLOCK_SIZE) {
		setBlockSize(...);
	}
}
\end{lstlisting}

\subsubsection{Feedback from \texttt{sat4j-core}}

We contacted the \texttt{sat4j-core} lead developer about two $void$ methods. One of them, \rev{named \texttt{removeConstr}\footnote{\url{https://gitlab.ow2.org/sat4j/sat4j/blob/09e9173e400ea6c1794354ca54c36607c53391ff/org.sat4j.core/src/main/java/org/sat4j/tools/xplain/Xplain.java\#L214}}}, was covered directly by only one test case to target a specific bug and avoid regression issues. The other method, \rev{named \texttt{learn}\footnote{\url{https://gitlab.ow2.org/sat4j/sat4j/blob/09e9173e400ea6c1794354ca54c36607c53391ff/org.sat4j.core/src/main/java/org/sat4j/minisat/core/Solver.java\#L384}}}, was covered indirectly by 68 different test cases. The lead developer considered the first method as helpful to realize that more assertions were needed in the covering test case. Consequently, he made one commit\footnote{\url{https://gitlab.ow2.org/sat4j/sat4j/commit/afab137a4c1a54219f3990713b4647ff84b8bfea}} to verify the behavior of this method.

The second \pseudotested{} method  was considered a bigger problem, because it implements certain key optimizations for better performance. \rev{The tests cases triggered the optimization code but did not leverage the optimized result. Their result were the same with or without the optimization code. Consequently, the developer made a new commit\footnote{\url{https://gitlab.ow2.org/sat4j/sat4j/commit/46291e4d15a654477bd17b0ce905926d24e042ca}} with an additional, more complex, test case where the advantages of the optimization could be witnessed.}

\subsubsection{Discussion}

We now discuss the main findings of this qualitative user study.

First, all developers agreed that it is easy to understand the problems identified by \pseudotested{} methods.
This confirms the fact that, we, as outsiders to those projects, with no knowledge or experience, can also grasp the issue and propose a solution. The developers acknowledged the relevance of the uncovered flaws.

Second, when developers were given the solution for free (through pull requests written by us), they accepted the test improvement. 

Third, when the developers were only given the problem, they did not always act by improving the test suite.
They considered that \pseudotested{} methods provide relevant information, and that it would make sense to enhance their test suites to tackle the issues. But they do not consider these improvements as a priority.
With limited resources, the efforts are directed to the development of new features and to fix existing bugs, not to improve existing tests.

Of the eight testing issues found, seven can be linked to oracle issues and one to an input problem.

\takeaway{Pseudo-tested methods uncover flaws in the test suite which are considered relevant by developers. These methods enable one to well understand the problem in a short time. However, fixing the test flaws requires some time and effort that cannot always be given, due to higher priority tasks such as new features and bug fixing.}

\subsection{RQ4: Which \pseudotested{} methods do developers consider worth an additional testing action?}
\label{sec:rq4}

To answer this question we contact the development teams directly. We select three projects for which  the developers have accepted to discuss with us: \texttt{authzforce}, \texttt{sat4j-core} and \texttt{spoon}. We set up a video call with the head of each development team. The goal of the call is to present and discuss a selection of \pseudotested{} methods in approximately 90 minutes. With this discussion, we seek to know which \pseudotested{} methods developers consider relevant enough to trigger additional work on the test suite and approximate their ratio on each project.

For projects \texttt{sat4j-core} and \texttt{spoon}, we randomly choose \~25\% of all \pseudotested{} methods. The third project, \texttt{authzforce}, has only 13 of such methods so we consider them all, as it is a number that can be discussed in reasonable time. We prepared a report for the developers that contains the list of \pseudotested{} methods, with the extreme transformations that were applied and the test cases covering the method. To facilitate the discussion we also included links to the exact version of the code we analyzed. This information was made available to the developers before the meeting.

For each method, we asked the developers to determine if: 1) given the structure of the method, and, 2) given its role in the code base, they consider it is worth spending time creating new test cases or fixing existing ones to specify those methods. We also asked them to explain the reasons behind their decision.

Table \ref{tab:todevelopers} shows the projects involved, footnotes with links to the online summary of the interviews, the number of \pseudotested{} methods included in the random sample, the number of methods worth an additional testing action and the percentage they represent with respect to the sample. We also show how much time we spent in the discussion.

\begin{table}
    \caption{The \pseudotested{} methods systematically analyzed by the lead developers, through a video call.}
	\label{tab:todevelopers}
	\centering
    \begin{tabular}{lrrrrrr}
        \toprule
        Project             & Sample size & Worth & Percentage  & Time spent (HH:MM)  \\
        \midrule
        \texttt{authzforce}\tablefootnote{\url{https://github.com/STAMP-project/descartes-experiments/blob/master/actionable-hints/authzforce-core/sample.md}} &  13 (100\%) &                    6 &        46\% & 29 min              \\  
		\texttt{sat4j-core}\tablefootnote{\url{https://github.com/STAMP-project/descartes-experiments/blob/master/actionable-hints/sat4j-core/sample.md}} &  35  (25\%) &                    8 &        23\% & 1 hr 38 min         \\
		\texttt{spoon}\tablefootnote{\url{https://github.com/STAMP-project/descartes-experiments/blob/master/actionable-hints/spoon/sample.md}}      &  53  (25\%) &                   16 &        23\% & 1 hr 14 min         \\
		\midrule
		Total               & 101         &                   30 &        30\% & 3 hr 21 min         \\
        \bottomrule
    \end{tabular}
\end{table}

We observe that only 23\% of \pseudotested{} methods in \texttt{sat4j-core} and \texttt{spoon}, the two largest projects, are worth additional testing actions (having the same percentage is purely coincidental). For \texttt{authzforce} the percentage of methods to be specified is 46\%, but the absolute number (6) does not differ much for \texttt{sat4j-core} (8). This indicates that, potentially, many \pseudotested{} come from functionalities considered less important or not a priority, therefore not well tested. 
The proportion of methods considered as worthless additional testing appears surprisingly high. It is important to notice that, among \pseudotested{} methods, developers find cases, in their own words, ``surprising'' and ``definitively not well tested, but they should be''. Even for the cases they don't consider important, a developer from \texttt{sat4j-core} state that they ``would like to know in which scenarios the transformation was discovered''.

We now enumerate the main reasons given by developers to consider a method worth or worthless spending time creating specific testing actions.
We also include the projects in which these reason manifested.

\textbf{Worthless specifying:} A \pseudotested{} method could be considered as useless to test, \ie, \emph{not important}, if it meets one of the following criteria:

\begin{itemize}
	\label{enum:reasons-unworthy}
	\item The code has been automatically generated (\texttt{spoon}).
	\item The method is part of debug functionalities, \ie, formatting a debug or log message or creating and returning an exception message or an exception object (\texttt{authzforce}).	
	\item The method is part of features that are not widely used in the code base or in external client code (\texttt{authzforce}, \texttt{sat4j-core}, \texttt{spoon}).
	\item The method has not been deprecated but its functionality is being migrated to another interface (\texttt{spoon}).
	\item The code of the method is considered as simple or trivial to need a specific test case (\texttt{sat4j-core}, \texttt{spoon}). Short methods, involving a few simple instructions are generally not considered worth to be specified by a direct unit test cases (\texttt{spoon}). Listing \ref{list:simple} shows two examples that \texttt{spoon} developers consider too simple to to be worth of additional testing actions. 
	\item Methods created just to complete an interface implementation (\texttt{spoon}). The object oriented design may involve classes that need to implement a given interface but do not actually need to provide a behavior for all methods. In those cases, developers write a placeholder body which they are not interested in testing.
	\item Receiving methods in a delegation pattern that add little or no logic when invoking the delegate (\texttt{spoon}). The delegation pattern exposes a method that simply calls another method (delegate). Delegate methods may have the same signature as the receiving method. The receiving method usually adds no or very little custom logic (e.g., provide a default value for unused parameters or process the returning value). Listing \ref{list:delegates} shows an example of this pattern that developers do not consider to be worth of additional testing actions. If the delegate is \pseudotested{} then the receiving method will be \pseudotested{} as well. The opposite does not have to be necessarily true. In any case, the method exposing the actual functionality to be tested is the delegate. The receiving method may not have the same importance. 
\end{itemize}

	\begin{lstlisting}[caption=Pseudo-tested method involving a delegation pattern., captionpos=b, label=list:delegates,float]
	...
	public void addArrayReference(CtArrayTypeReference<?> typeReference) {
	    arrayTypeReference.setComponentType(typeReference);
	}
	...
	\end{lstlisting}

	\begin{lstlisting}[caption=Pseudo-tested methods considered as too simple to require more testing actions., captionpos=b, label=list:simple,float]
		...
		public void externalState() {
			this.selectedState = external;
		}
		...
		
		public boolean matches(CtElement e) {
			e.setFactory(f);
			return false;
		}
	\end{lstlisting}

\textbf{Worth specifying with additional tests:}	On the other hand, developers provided the following reasons when they consider a  \pseudotested{} method to be worth of additional testing actions: 
	
\begin{itemize}
	\label{enum:reasons-worthy}
	\item A method that supports a core functionality of the project or part of the main responsibility of the declaring class (\texttt{authzforce}, \texttt{sat4j-core}, \texttt{spoon}). For example, we find a class named \texttt{VisitorPartialEvaluator} which implements a visitor pattern over a Java program Abstract Syntax Tree (AST). This class simplifies the AST by evaluating all expressions that could be statically reduced. The method \texttt{visitCtAssignment}\footnote{\url{https://github.com/INRIA/spoon/blob/fd878bc71b73fc1da82356eaa6578f760c70f0de/src/main/java/spoon/support/reflect/eval/VisitorPartialEvaluator.java\#L515}}, declared in this class, handles assignment instructions and was found to be \pseudotested{}. Assignments may influence the evaluation result, so this method plays an important role in the class.
	\item A method supporting a functionality that is widely used in the code base. It could be the method itself that is being frequently used or the class that declares the method (\texttt{authzforce}, \texttt{sat4j-core}, \texttt{spoon}).
	\item A method known to be relevant for external client code (\texttt{sat4j-core}).
	\item A new feature that is partially supported or not completed yet, which requires a clear specification (\texttt{spoon}).
	\item Methods which are the only possible way to access certain features of a class (\texttt{authzforce}). For example, a public method that calls several private methods which actually contain the implementation of the public behavior.
	\item Method verifying preconditions (\texttt{authzforce}). These methods guarantee the integrity of the operations to be performed. Listing \ref{list:precondition} shows an example of one of those methods considered to be important to specify. Despite the simplicity of the implementation, the \texttt{authzforce} developers consider that it is important to  specify them as accurately as possible.
\end{itemize}

\begin{lstlisting}[caption=A simple method that checks a precondition, captionpos=b, label=list:precondition,float]
	protected final void checkNumberOfArgs(final int numInputs)
	{
		if (numInputs != 3)
		{
			throw new IllegalArgumentException(...);
		}
	}
\end{lstlisting}

We have observed cases where a method meets criteria to be worth of specification and at the same time to be worthless of additional testing actions. The final decision of developers in those cases is subjective and responds to their internal knowledge about the code base. This means that it is difficult to devise an automatic procedure able to automatically determine which methods are worth of additional testing actions.

\takeaway{In a sample of 101 \pseudotested{} methods, systematically analyzed by the lead developers of 3 mature projects, 30 methods (30\%) were considered worth of additional testing actions. The developer decisions are based on a deep understanding of the application domain and design of the application. This means that it is not reasonable to prescribe the absolute absence (zero) of \pseudotested{} methods.}

\section{Threats to validity}
\label{sec:threats}

\textbf{RQ1 and RQ2}. 
A threat to the quantitative analysis of RQ1 and RQ2 relates to external validity: 
\begin{itemize}

\item Some extreme transformations could generate programs that are equivalent to the original. Given the nature of these transformations many of possible equivalent variants are detected by inspecting the method before applying the transformation. Methods with empty body and those returning a constant value, are skipped from the analysis. This is a problem extreme transformations have in common with traditional mutation testing. \rev{The equivalent mutant problem could also affect the value of the mutation scores computed to answer RQ2.}

\item The values used to transform the body of non-void and non-boolean methods may affect their categorization as \pseudotested{} or required. A different set of values may produce a different categorization. Since only one detected value is needed to label a method as required, then we actually produce an over-estimation of these methods. More values could be used to reduce the final set. In our study we find only 916 of non-void and non-boolean \pseudotested{} methods which represents a 36\% of the total number of methods. $void$ and $boolean$ methods tend to produce more \pseudotested{} methods in our study subjects.

\item As pointed by Goran and Ivankovic \cite{goran2018state}, mutation testing results can be affected by the programming language. This affects both, the extreme transformations performed and the mutation testing validation in RQ2. All our study subjects are \java{} projects, so our findings can not be generalized to other languages.

\item \rev{We have considered all test cases equally and did not attempt to distinguish between unit and integration tests. We made this decision because such a distinction would require setting an arbitrary threshold above which a JUnit test case is considered an integration test. Yet, considering all test cases equally could influence the amount of \pseudotested{} methods.}
\end{itemize}

\textbf{RQ3 and RQ4}. 
The outcome of the qualitative analysis is influenced by our insight into each project, the insight of the developers consulted, the characteristics of each code base and the methods presented. Some of the teams showed more interest and gave more importance to the findings than others. This was expected. Not all developers had strong opinions regarding the presented issues.	

\section{Related work}
\label{sec:related}

Our work is inspired by the original paper on  \pseudotested{} methods \cite{niedermayr_will_2016}. The authors aim at assessing the relevance of test coverage with respect to test adequacy.  They introduce the concept of \pseudotested{} method to analyze the exact set of methods, which code is covered and which behavior is poorly assessed by the test suite. They also study the type of test involved with \pseudotested{} methods, and aim at  answering the question of whether code coverage is an indicator of test quality. 

The novelty of our contribution with respect to this paper is three-fold. First, we perform a study with novel study objects (19 among the 21 projects studied here are not analyzed by Niedermayr and colleagues in this paper) and a different tool to detect the \pseudotested{} methods. This mitigates both internal and external threats to the validity of Niedermayr's results.
Second, we perform a  novel study about the adequacy of the test suite for \pseudotested{} methods. Third, our most significant novel contribution consists in extensive exchanges and interactions with software developers to understand the type of testing issues that are revealed by \pseudotested{} methods, as well as the characteristic of \pseudotested{} that developers consider worth an additional testing effort.

As discussed in RQ3, \pseudotested{} methods reveal weaknesses in the oracles of the test suite. Previous works have devised techniques to assess the quality of the test oracle. Schuler and Zeller \cite{schuler2013checked} introduce the concept of checked coverage, as the percentage of program statements that are executed by the test suite and whose effects are also checked in the oracles. They compare this metric to code coverage and mutation testing with respect to their ability at assessing oracle decay. They perform manual checks on seven real software projects and conclude that checked coverage is more realistic than the usual coverage. 

Jahangirova et al. \cite{jahangirova2016test} propose a technique for assessing and improving test oracles. They use mutation testing to increase the fault detection capabilities of the oracles and automated unit test generation to reduce the number correct executions rejected by the assertions. Their approach is shown to be effective in five real software projects. The fault detection ratio, approximated with the mutation score, is increased by 48.6\% in average.

Staats et al. \cite{staats2011programs} extend the work of Gourlay \cite{gourlay1983mathematical} to provide a mathematical framework to capture how oracles interrelate with specifications, programs and tests. In this proposal an oracle is defined as a predicate whose domain is the product of the set of tests and the set of programs. Formal definitions of oracle completeness and soundness are given. The authors provide hints to the oracle selection problem and express criteria to compare oracles. The authors revisit concepts such as coverage and mutation testing under the light of their proposal.

The work of Androutsopoulos et al. \cite{androutsopoulos2014analysis} focuses on how faulty program states propagate to observable points in the program. They observe that one in ten test inputs fail to propagate to observable points. The authors provide an information theoretic formulation of the phenomenon through five metrics and experiment with 30 programs and more than 7M test cases. They state that better understanding the causes of failed error propagation leads to better testing.

Mutation testing is a well known technique to assess the fault detection capabilities of the test suite \cite{demillo1978hints}. The traditional approach has been evaluated against real faults in several occasions \cite{daran_software_1996,andrews_is_2005,just_are_2014}. The evidence presented supports that mutation testing is able to create effective program transformations under the assumption that programming errors are generally small and complex faults can be detected by tests which also detect simpler issues. However, some concerns has been raised regarding the conditions for those assumptions to be true \cite{gopinath_mutations_2014}.

Several works perform comparative analyses of mutation tools and confirm our choice of \pit{} for mutation analysis.
Delahaye and Bousquet \cite{delahaye_comparison_2013}, then Kintis et al. \cite{kintis_analysing_2016}  compare mutation tools from the usability point of view concluding that \pit{} is one of the best alternatives for concurrent execution and adaptability to distinct requirements.  In a follow-up paper, Laurent and colleagues \cite{laurent_assessing_2017}  propose to improve \pit{} with an extended set of mutation operators shown to obtain better results. 

More recently, Gopinath et al. \cite{gopinath_does_2017} performed a comprehensive analysis of  3 software tools, including \pit{}, with 27 projects. They run several statistical analyses to compare the performance of the tools considering projects and tools characteristics against raw mutation score, a refined mutation score to mitigate the impact of equivalent mutants and the relationship among mutants. They conclude that \pit{} is slightly better than the other tools and that the specific project characteristics have a high impact on the effectiveness of the mutation analysis. 

\rev{Other mutation tools, such as Major \cite{JustSK2011}, could be extended as well to implement extreme transformations and detect \pseudotested{} methods.}

\rev{Several works from the mutation testing literature  propose to transform programs using only deletion of statements, blocks, variables, operators and constants \cite{delamaro2014designing,deng2013empirical,untch2009reduced}. These approaches create much fewer mutants than traditional mutation operators showing decreases below 80\%, but still remain at the instruction level which can difficult their understanding. Durelli et al. \cite{durelli2017are} actually state that these mutants are as time-consuming as the traditional operators when developers try to determine if they are equivalent to the original code. None of these works use  extreme transformations as we have done here. Our work and our our exchanges with developers show that these transformations that delete the complete method body  are easier to understand.}

\section{Conclusion}
\label{sec:conclusion}

In this work, we provide an in-depth analysis of \pseudotested{} methods found in open source projects. These methods, first coined by \theoriginalauthors, are intriguing from the perspective of the interaction between a program and its test suite: their code is executed when running the test suite, yet, none of their effects are assessed by the test cases.

Our key findings are as intriguing as the original concept. 
First, we observe that all 21 mature Java projects that we study include such methods: from 1\% to 46\% in our dataset. Second, we confirm that \pseudotested{} methods are poorly tested, compared to the required methods: the mutation score of the former is systematically and significantly lower than the score of the latter. 
Third, our in-depth qualitative analysis of \pseudotested{} methods and feedback from developers reveals the following facts:
\begin{itemize}
	\item We assessed the relevance of  \pseudotested{} methods as concrete hints to reveal weak test oracles. These issues in the suite were confirmed by the developers, who accepted the pull requests that we proposed, to fix weak oracles.
	\item Less than 30\% of \pseudotested{} methods in a sample of 101 represent an actual hint for further actions to improve the test suite. Among the 70\% considered as worthless an additional testing action we found methods not widely used in the code base, automatically generated code, trivial methods, helper methods for debugging and receiving methods in delegation patterns.
	\item The \pseudotested{} methods that actually reveal an issue in the interaction between the program and its test suite are involved in core functionalities, are widely used in the code base, are used by external clients or verify preconditions on input data.
\end{itemize}

In the light of these conclusions, the immediate next step in our research agenda is to investigate an automatic test generation technique targeted towards \pseudotested{} methods. This technique shall kill two birds with one stone: improve the adequacy of the test suite for \pseudotested{} methods; let the developers focus their efforts on core features and relieve them from the test improvement task. \rev{The current state of \pit{} and Descartes allows to exclude methods from the analysis based on the name, on some structural patterns, or considering the annotations marking the method. These filters are not able to express some of the criteria exposed in Section \ref{sec:rq4}, for example, methods which are not widely used in the codebase. A future line will consist in improving the static analysis capacities of the tools to reduce even more the number of methods that developers do not consider worhty for testing. Further steps in the characterization of \pseudotested{} methods could include the comparison of the statement coverage ratio between \pseudotested{} and required methods, in the same way we have done considering the method-level mutation score. Another future task could be to assess the fault detection capabilities of extreme transformations in the same way it has been done for mutation operators \cite{gopinath_mutations_2014}.}

\section*{Acknowledgement}

We would like to acknowledge the invaluable help and feedback provided by the 
development teams of \texttt{authzforce}, \texttt{spoon} and \texttt{sat4j-core}.
We also express our appreciation to Simon Urli, Daniel Le Berre, 
Arnaud Blouin, Marko Ivankovi\'c, Goran Petrovic and Andy Zaidman
for their feedback and their very accurate suggestions.
This work has been partially supported by the EU Project STAMP ICT-16-10 No.731529 and by the Wallenberg Autonomous Systems and Software Program (WASP) funded by the Knut and Alice Wallenberg Foundation.

\bibliographystyle{abbrv}
\bibliography{bibliography}

\clearpage
\appendix
\section{Source code for the study subjects}
\label{app:source}

\textbf{Legend.} Column ``Project" list the projects included in the present study. 
Column ``URL" contains links to the available source code.
Column ``Commit ID" contains the SHA-1 hash identifying the commit with the source code state that was used in this study. 

\begin{table}[ht]
    \begin{adjustbox}{max width=\textwidth,center=\textwidth}
	\begin{tabular}{llll}
		\toprule
		Project & URL/Commit ID\\
        \midrule
        \multirow{2}{*}{\texttt{authzforce}} & \url{https://github.com/authzforce/core.git} \\
        & 81ae56671bc343eabf2bc99ee0c51ba6ae28d649 \\[0.1cm]
        \multirow{2}{*}{\texttt{aws-sdk-java}}        & \url{https://github.com/aws/aws-sdk-java} \\ 
        & b5ae6ce44f4b5053a9a0255c9648f3073fafcf55 \\[0.1cm]
        \multirow{2}{*}{\texttt{commons-cli}}         & \url{https://github.com/apache/commons-cli} \\ 
        & c246bd419ee0efccd9a96f9d33486617d5d38a56 \\[0.1cm]
        \multirow{2}{*}{\texttt{commons-codec}}       & \url{https://github.com/apache/commons-codec} \\ 
        & e9da3d16ae67f2940a0bbdf982ecec19a0481981 \\[0.1cm]
        \multirow{2}{*}{\texttt{commons-collections}}   & \url{https://github.com/apache/commons-collections} \\ 
        & db189926f7415b9866e76cd8123e40c09c1cc67e \\[0.1cm]
        \multirow{2}{*}{\texttt{commons-io}}          & \url{https://github.com/apache/commons-io} \\ 
        & e36d53170875d26d59ca94bd376bf40bc5690ee6 \\[0.1cm]
        \multirow{2}{*}{\texttt{commons-lang}}        & \url{https://github.com/apache/commons-lang} \\ 
        & e8f924f51be5bc8bcd583ea96e5ef25f9b2ca72a \\[0.1cm]
        \multirow{2}{*}{\texttt{flink-core}}          & \url{https://github.com/apache/flink/tree/master/flink-core} \\ 
        & 740f711c4ec9c4b7cdefd01c9f64857c345a68a1 \\[0.1cm]
        \multirow{2}{*}{\texttt{gson}}                & \url{https://github.com/google/gson} \\ 
        & c3d17e39f1cb6ec41496e639ab42f7e7cca3b465 \\[0.1cm]
        \multirow{2}{*}{\texttt{jaxen}}               & \url{https://github.com/jaxen-xpath/jaxen} \\ 
        & a8bd80599fd4d1c9aa1248d3276198535a30bfc5 \\[0.1cm]
        \multirow{2}{*}{\texttt{jfreechart}}          & \url{https://github.com/jfree/jfreechart} \\ 
        & a7156d4595ff7f6a7c8dac50625295c284b86732 \\[0.1cm]
        \multirow{2}{*}{\texttt{jgit}}                & \url{https://github.com/eclipse/jgit} \\ 
        & 1513a5632dcaf8c6e2d6998427087e11ba35566d \\[0.1cm]
        \multirow{2}{*}{\texttt{joda-time}}           & \url{https://github.com/JodaOrg/joda-time} \\ 
        & 6ad133837a4c4f8199d00a05c3c16267dbf6deb8 \\[0.1cm]
        \multirow{2}{*}{\texttt{jopt-simple}}         & \url{https://github.com/jopt-simple/jopt-simple} \\ 
        & b38b70d1e7685766ab400d8b57ef9ca9c010e0bb \\[0.1cm]
        \multirow{2}{*}{\texttt{jsoup}}               & \url{https://github.com/jhy/jsoup} \\ 
        & 35e80a779b7908ddcd41a6a7df5f21b30bf999d2 \\[0.1cm]
        \multirow{2}{*}{\texttt{pdfbox}}          & \url{https://github.com/apache/pdfbox} \\ 
        & 09e9173e400ea6c1794354ca54c36607c53391ff \\[0.1cm]
        \multirow{2}{*}{\texttt{sat4j-core}}              & \url{https://gitlab.ow2.org/sat4j/sat4j/tree/master/org.sat4j.core} \\ 
        & 1a0127645bf98b768ee3628076d0246596dd15eb \\[0.1cm]
        \multirow{2}{*}{\texttt{scifio}}              & \url{https://github.com/scifio/scifio} \\ 
        & 2760af6982ad18aab400e9cd99b9f63ef2495333 \\[0.1cm]
        \multirow{2}{*}{\texttt{spoon}}               & \url{https://github.com/INRIA/spoon} \\ 
        & fd878bc71b73fc1da82356eaa6578f760c70f0de \\[0.1cm]
        \multirow{2}{*}{\texttt{urbanairship}}        & \url{https://github.com/urbanairship/java-library} \\ 
        & aafc049cc1cd3971c62a3dfc1d72cfe61160f32c \\[0.1cm]
        \multirow{2}{*}{\texttt{xwiki-rendering}}     & \url{https://github.com/xwiki/xwiki-rendering} \\ 
        & cb3c444fb743e073eefbac2b44351a6166d94ac1 \\
		\bottomrule
    \end{tabular}
\end{adjustbox}
\end{table}

\end{document}